\documentclass[journal]{IEEEtran}
\usepackage{amsmath,amsfonts}
\usepackage{algorithmic}
\usepackage{algorithm}
\usepackage{array}
\usepackage[caption=false,font=normalsize,labelfont=sf,textfont=sf]{subfig}
\usepackage{textcomp}
\usepackage{stfloats}
\usepackage{url}
\usepackage{verbatim}
\usepackage[table]{xcolor}
\usepackage{graphicx}
\usepackage{cite}
\usepackage{booktabs}
\usepackage{multirow}
\usepackage{algorithm}
\usepackage{algorithmic}
\usepackage[pagebackref=true,breaklinks=true,colorlinks,bookmarks=false]{hyperref}
\definecolor{mycolor}{RGB}{198,234,251}
\definecolor{unitcolor}{RGB}{235,235,235}

\definecolor{mycolor1}{RGB}{62,128,128}
\definecolor{mycolor2}{RGB}{30,18,0}

\makeatletter
\newcommand{\spthickhline}{%
    \noalign {\ifnum 0=`}\fi \hrule height 1.5pt
    \futurelet \reserved@a \@xhline
}

\begin{document}

\title{Semi-supervised Semantic Segmentation with Multi-Constraint Consistency Learning}

\author
{
Jianjian Yin, Tao Chen, Gensheng Pei, Huafeng Liu, Yazhou Yao, Liqiang Nie, Xiansheng Hua

\thanks{J.~Yin, T.~Chen, G.~Pei, H.~Liu, Y.~Yao are with the School of Computer Science and Engineering, Nanjing University of Science and Technology, Nanjing, China. Co-corresponding author: Yazhou Yao and Tao Chen.}% 

\thanks{L.~Nie is with the School of Computer Science and Technology, Harbin Institute of Technology (Shenzhen), Shenzhen, China.}

\thanks{X.~Hua is with the Terminus Group, Beijing, China.}
}

\maketitle

\begin{abstract}
Consistency regularization has prevailed in semi-supervised semantic segmentation and achieved promising performance. However, existing methods typically concentrate on enhancing the Image-augmentation based Prediction consistency and optimizing the segmentation network as a whole, resulting in insufficient utilization of potential supervisory information. In this paper, we propose a \textbf{M}ulti-\textbf{C}onstraint  \textbf{C}onsistency \textbf{L}earning (\textbf{MCCL}) approach to facilitate the staged enhancement of the encoder and decoder. Specifically, we first design a feature knowledge alignment (FKA) strategy to promote the feature consistency learning of the encoder from image-augmentation. Our FKA encourages the encoder to derive consistent features for strongly and weakly augmented views from the perspectives of point-to-point alignment and prototype-based intra-class compactness. Moreover, we propose a self-adaptive intervention (SAI) module to increase the discrepancy of aligned intermediate feature representations, promoting Feature-perturbation based Prediction consistency learning. Self-adaptive feature masking and noise injection are designed in an instance-specific manner to perturb the features for robust learning of the decoder.  Experimental results on Pascal VOC2012 and Cityscapes datasets demonstrate that our proposed MCCL achieves new state-of-the-art performance. The source code and models are made  available at \url{https://github.com/NUST-Machine-Intelligence-Laboratory/MCCL}. 
\end{abstract}


\section{Introduction}
\IEEEPARstart{R}{ecent} substantial advancements in deep neural networks \cite{chen2022saliency,cai2024poly,pei2023hierarchical,pei2024videomac,tangcsvt,chen2021semantically,yao2021non,li2022weakly,pei2022hierarchical,zhoubocvpr,genshengcvpr} have significantly improved the performance of various tasks.  Semantic segmentation \cite{chen2023multi,pei2023hierarchical,long2015fully,zhao2017pyramid,chen2024knowledge,chen2024spatial,liu2023fecanet}, a prevalent task in computer vision, drives the model to interpret and analyze scenes. Its primary goal is to allocate specific class labels to each pixel in an image.
Fully supervised semantic segmentation methods \cite{long2015fully,nie2022mign,li2022deep,xie2021segformer,cheng2021per} mainly focus on enhancing the performance of semantic segmentation by designing novel model architectures and modeling basic pixel features to enhance the feature representation of pixels. However, a major challenge lies in the need for comprehensive pixel-level annotations for training, which are labor-intensive and difficult to acquire. To mitigate this, semi-supervised semantic segmentation approaches have been introduced. These methods utilize a combination of limited labeled data and a substantial amount of unlabeled data, as proposed in various studies \cite{alonso2021semi,chen2021semi,mendel2020semi}.

\begin{figure}[t]
	\centering
	\includegraphics[width=\linewidth]{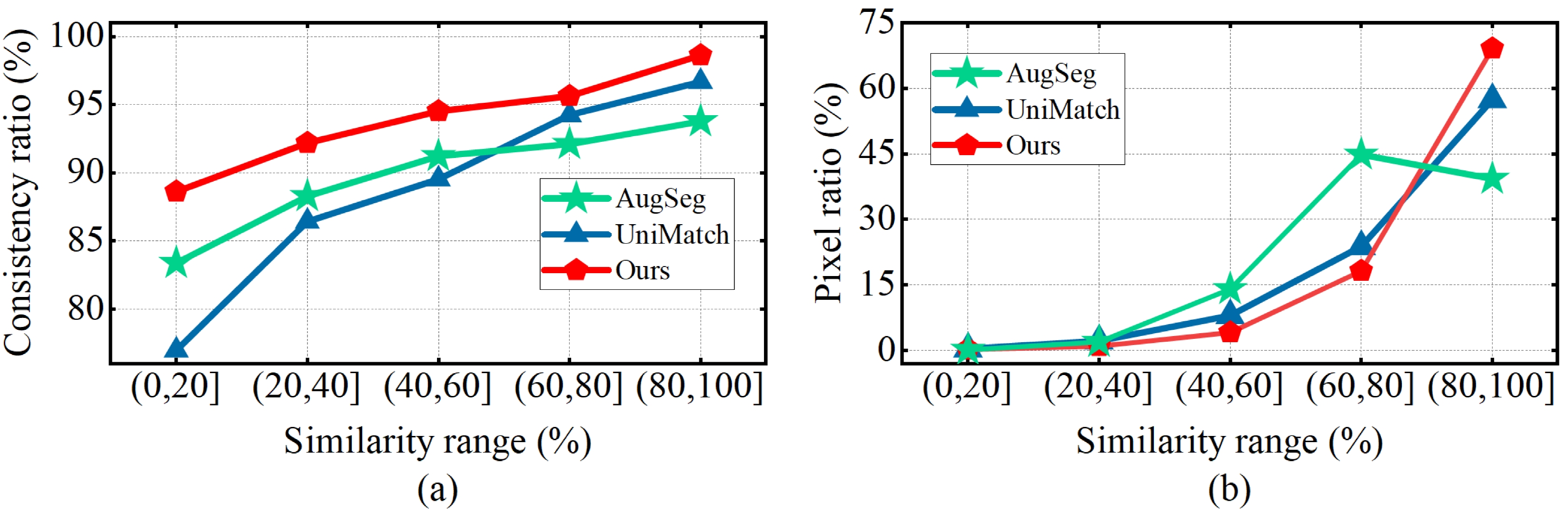}
	\caption{Explanation of motivation. (a) The comparison of consistency ratio with previous SOTA methods across different feature similarity ranges between Strongly and Weakly augmented View (SWV) features.  Considering the higher similarity of SWV features leads to better prediction consistency, we propose to further enhance the consistency learning with Image-augmentation based Feature (IF) consistency to generate  more similar SWV features. (b) The comparison of pixel ratio across different feature similarity ranges. Our proposed feature knowledge alignment (FKA) strategy effectively increases the ratio of pixels with higher-similarity SWV features.}
	\label{fig_first:movtivation}
\end{figure}

Semi-supervised semantic segmentation aims to utilize the insights gained from labeled data to extract more valuable information from unlabeled data, a key focus of this study. The learning framework of generative adversarial networks \cite{goodfellow2014generative,mittal2019semi,souly2017semi} was initially applied in the realm of semi-supervised semantic segmentation and has demonstrated advancement. 
Subsequently, Consistency regularization \cite{kim2020structured,ouali2020semi,ke2020guided,wang2023conflict,yuan2023}, a prominent technique in this field, strives to ensure that the network produces uniform predictions for variously augmented versions of the same input. For instance, several studies \cite{feng2022dmt,tarvainen2017mean,kim2022conmatch} have concentrated on  enhancing prediction consistency by introducing different perturbations to the model. Concurrently, various data augmentation strategies \cite{yang2023revisiting,zhao2023augmentation,Geoffrey,sohn2020fixmatch,olsson2021classmix} have been developed to enhance the generalization capabilities of the network. Nonetheless, current methods in consistency regularization primarily focus on Image-augmentation based Prediction (IP) consistency  and the overall optimization of the segmentation network, which often leads to insufficient utilization of potential supervisory information, e.g. feature consistency.

We perform an in-depth analysis of the features in strongly and weakly augmented views using the state-of-the-art consistency regularization methods, UniMatch \cite{yang2023revisiting} and AugSeg \cite{zhao2023augmentation}.  We normalize the similarity between pixel features of strongly and weakly augmented views. 
The fundamental principle of consistency regularization is to maximize consistency prediction between  different augmented views, while Fig. \ref{fig_first:movtivation} (a)  illustrates that the consistency of prediction improves with increasing similarity. Consequently, one of the primary focuses of this paper is to promote feature consistency learning within the network.

\begin{figure}[t]
	\centering
	\includegraphics[width=\linewidth]{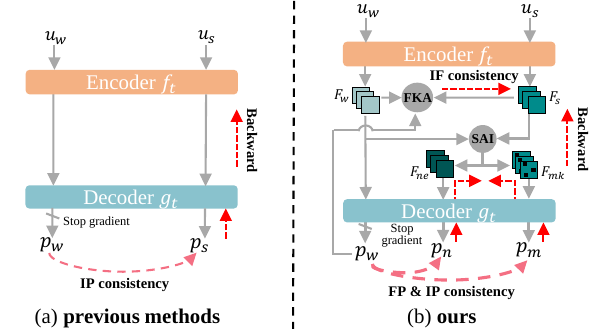}
	\caption{Comparison with previous methods. (a) The existing methods primarily emphasize the Image-augmentation based Prediction (IP) consistency. (b) In contrast, our approach differs by focusing more on  Image-augmentation based Feature (IF) consistency and  Feature-perturbation based Prediction (FP) consistency. FKA encourages the encoder to perform consistency learning on the features ($F_s$ and $F_w$) of strongly and weakly augmented views ($u_s$ and $u_w$). SAI adaptively intervenes in $F_s$  to generate specific perturbed features ($F_{ne}$ and $F_{mk}$). The predictions (
		$p_n$ and $p_m$) of $F_{ne}$ and $F_{mk}$ are supervised by weakly augmented view prediction $p_w$ to achieve FP and IP consistency.}
	\label{fig1:movtivation}
	
\end{figure}

Building on the observations presented above, this paper proposes a Multi-Constraint  Consistency Learning  approach to facilitate the staged enhancement of the encoder and decoder by imposing additional consistency constraints, thus fully utilizing the potential supervisory information in the network. As shown in Fig. \ref{fig1:movtivation}, previous methods have overlooked the potential of Image-augmentation based Feature (IF) consistency. A robust encoder should yield consistent features for both strongly and weakly augmented views. To achieve this, we propose a feature knowledge alignment (FKA) strategy, enhancing IF consistency learning in the encoder. 
In particular, strong augmentation, applied to the original images, such as color jitter, graying, and Gaussian blur, introduces complex contextual information to the strongly augmented views. In contrast, weak augmentation, limited to horizontal and vertical flipping, has a negligible effect on contextual information. 
Building upon it, we encourage features from weakly augmented views to guide those from strongly augmented views, utilizing both  point-to-point alignment and prototype-based intra-class feature compactness.
Point-to-point alignment fundamentally relies on using the similarity between features at corresponding positions in different augmented views as a loss function. This encourages features in the strongly augmented feature map to align with those in the weakly augmented one, thereby allowing the network to extract richer knowledge from complex contextual information.
However, the network may still misclassify some outlier features that are far from the cluster centers during training. Therefore, we resort to prototype-based intra-class feature compactness. After obtaining the feature prototype for each class, we select several intra-class features from the weakly augmented view that are most similar to the class prototypes. Simultaneously, we identify outlier features from the strongly augmented view that are more distant from the prototypes. 
Then, we minimize the nearest neighbor similarity loss between these selections to encourage the network to focus more on outlier features.

Fig. \ref{fig_first:movtivation} (b)  illustrates that the FKA strategy substantially enhances feature consistency learning for the encoder.  The high degree of similarity between features from the strongly and weakly augmented views results in consistent predictions from the decoder. While this outcome may seem desirable, it restricts the decoder's learning scope within the feature space. The strongly and weakly augmented view features used for decoder training are highly similar, which simplifies the training process and may result in the decoder becoming trapped in local optima. In the field of semi-supervised semantic segmentation, a robust and efficient decoder should maintain the ability to generate consistent features even when confronted with paired features that exhibit considerable differences. 
To address this, we design a self-adaptive intervention (SAI) module, designed to enhance decoder training through Feature-perturbation based Prediction (FP) consistency. The SAI module, tailored for each instance, acts on features of strongly augmented views, thereby broadening the decoder's learning capability in a wider feature space. 
The module comprises two components: self-adaptive feature masking and noise injection. The former masks highly activated regions in the feature map of strongly augmented views based on inter-view feature similarity, reducing network dependency on these regions and averting overfitting. The latter injects self-adaptive noise into features to diversify them, thus expanding the decoder's learning range. The outcomes of these interventions are guided by predictions from the weakly augmented view to achieve FP consistency. Importantly, as the gradient backpropagation between the encoder and decoder remains uninterrupted, the conventional Image-augmentation based Prediction (IP) consistency is also inherently sustained in our Multi-Constraint Consistency Learning. Our contributions can be summarized as follows: 

(1) We propose a Multi-Constraint Consistency Learning approach to facilitate the staged enhancement of the encoder and decoder by imposing additional consistency constraints on the network, maximizing the utilization of available supervisory information.

(2) A feature knowledge alignment strategy is designed to encourage the feature consistency learning of the encoder from image-augmentation,  specifically from the perspectives of point-to-point alignment and  prototype-based intra-class feature compactness.

(3) We design a self-adaptive intervention module to promote prediction consistency of feature-perturbation, enabling the decoder to learn more useful information from the broader feature space.

(4) Extensive experiments on the Pascal VOC2012 and Cityscapes datasets demonstrate that our proposed method achieves new state-of-the-art performance.

\section{Related Works}

\subsection{Semantic Segmentation}
Semantic segmentation aims to assign the correct label to every pixel in an image, allowing machines to analyze scenes and finds wide applications across fields including autonomous driving, medical image analysis, and augmented reality.  Prior semantic segmentation methods \cite{wang2020dual,wang2021efnet,zhou2022canet,wang2020deep} have concentrated on crafting innovative network architectures to achieve more robust pixel feature representations. Examples include PSPNet \cite{zhao2017pyramid}, integrating a feature pyramid module, and CCNet \cite{huang2019ccnet}, which employs cross-channel and bidirectional attention mechanisms. Current mainstream semantic segmentation methods \cite{yang2022cross, Jin_2021_ICCV, jin2022mcibi++,jin2024idrnet,zhou2024prototype} strive to extract diverse semantic information from image context to aid the decoder in classification, thus enhancing segmentation accuracy. MCIBI++ \cite{jin2022mcibi++} assumes that class semantic features at the dataset level follow a Gaussian distribution.  During each training and testing instance, semantic features for each class are randomly sampled from their respective mean and variance, thereby enriching the basic pixel feature representations.  IDRNet \cite{jin2024idrnet} guides contextual relationship modeling between pixels by deletion diagnostic procedure, thus reducing reliance on extensive prior information. PSS \cite{zhou2024prototype} uses a set of prototypes to describe the features of each class to construct a more discriminative feature embedding space. The exceptional performance exhibited by the methods mentioned above heavily relies on the quantity of available labeled data.

\subsection{Semi-supervised Semantic Segmentation}
Semi-supervised semantic segmentation (S4) endeavors to extract valuable information from unlabeled images, leveraging insights from a limited dataset of labeled images to improve model performance. Initially, the focus was predominantly on using generative adversarial networks (GANs) \cite{Wei,mittal2019semi,lee2023saliency,qi2019ke,souly2017semi} to yield high-quality predictions. However, recent advancements have been driven by pseudo-labeling methods \cite{chen2021semi,ibrahim2020semi,liang2023logic,zhou2021c3}, demonstrating significant progress. 
Unreliable pseudo labels are used by U$^2$PL \cite{wang2022semi} for contrastive learning to enhance the performance of the model. 
The LDR \cite{liang2023logic} approach capitalizes on relational knowledge between class semantic concepts to refine incorrect pseudo labels. FPL \cite{qiao2023fuzzy} promotes adaptive fuzzy positive predictions while concurrently minimizing the likelihood of false negatives. Lastly, ESL \cite{Ma_2023_ICCV} employs high-entropy predictions to dynamically preserve high-probability classes.  DCC \cite{Lai_2021_CVPR} focuses on context consistency in randomly sized local patches through contrastive learning. Differently, our proposed method emphasizes learning the consistency of paired pixel features in the global context using similarity loss.

Consistency regularization represents a significant research avenue in semi-supervised semantic segmentation, with recent methodologies \cite{peng2020deep,ijcai2022p226,Lai_2021_CVPR,Fan_2022_CVPR,miyato2018virtual} delivering commendable results. RC$^2$L \cite{ijcai2022p226} introduces region-level contrast along with consistency regularization, effectively reducing the impact of erroneous pixel noise on pixel-level regularization. PS-MT \cite{liu2022perturbed} innovates with an auxiliary teacher model and a more stringent confidence-weighted cross-entropy (CE) loss, replacing mean squared error loss to heighten segmentation accuracy on unlabeled data. MKD \cite{yuan2023} advocates for mutual knowledge distillation, using two auxiliary mean models to supervise and facilitate knowledge exchange between student models. UniMatch \cite{yang2023revisiting} devises a dual-stream augmentation approach, enabling the simultaneous supervision of two strongly augmented views by a weakly augmented view. 
AugSeg \cite{zhao2023augmentation} and Classmix \cite{olsson2021classmix} both aim to use a variety of data augmentation techniques to enhance prediction consistency in neural networks. However, these methods are primarily centered on Image-augmentation based Prediction (IP) consistency. In contrast, our approach emphasizes leveraging both Image-augmentation based Feature (IF) consistency and Feature-perturbation based Prediction (FP) consistency, by introducing additional consistency constraints to facilitate staged improvements in both the encoder and decoder.

\begin{figure*}[t]
\centering
\includegraphics[width=1\textwidth]{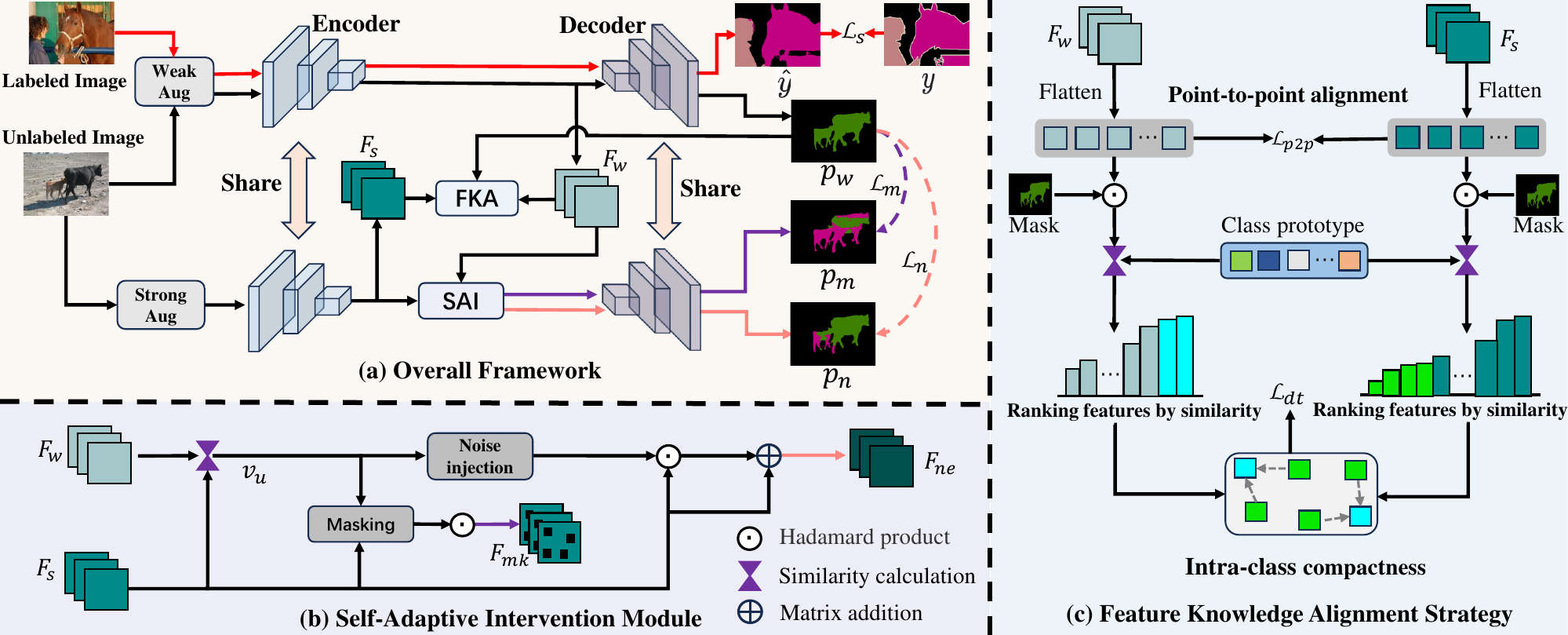}

\caption{The network architecture of Multi-Constraint Consistency Learning. The training process for labeled and unlabeled images is conducted simultaneously. The black area in $F_{mk}$ indicates that the area is filled with zero. Aug is the abbreviation for augmentation. } 
\label{fig2:Net}
\end{figure*}

\section{Proposed Method}
This section describes the content of our method in detail. Initially, we introduce the overall architecture and training process of the network, followed by a successive introduction of the feature knowledge alignment module and the self-adaptive intervention module. Finally, several loss functions are summarized for the optimization of network parameters.

\subsection{Overall Framework}
This paper investigates extracting more valuable information from unlabeled data, utilizing the insights gained from a limited labeled data denoted as ${S_l} = \{ ({x_i},{y_i})\} _{i = 1}^{|{S_l}|}$ and a vast collection of unlabeled data  ${S_u} = \{ ({u_i})\} _{i = 1}^{|{S_u}|}$. As depicted in Fig. \ref{fig2:Net} (a), a weakly augmented labeled image is fed into the network to produce the prediction $\hat{y}$, guided by the ground truth $y$.  In contrast, an unlabeled image first undergoes  strong and weak augmentations simultaneously, and is then input into the encoder, resulting in two different sets of feature representations, $F_s$ and $F_w$. Our Multi-Constraint Consistency Learning (MCCL) approach integrates a feature knowledge alignment strategy and a self-adaptive feature intervention module. The feature knowledge alignment strategy employs similarity measures to regulate the features of the strongly augmented view $F_s$, focusing on point-to-point alignment and prototype-based intra-class feature compactness. The self-adaptive feature intervention module generates self-adaptive masking feature representations $F_{mk}$ and noisy feature representations $F_{ne}$, based on the cosine similarity between $F_s$ and $F_w$,  tailored for each instance. Ultimately, the predictions from $F_{ne}$ and $F_{mk}$ are supervised by the prediction from the weakly augmented view $p_w$.

\subsection{Feature Knowledge Alignment}
Consistency regularization methods leverage prediction results from weakly augmented views to supervise those from strongly augmented views, thereby attaining Image-augmentation based Prediction (IP) consistency. However, these methods often neglect the crucial aspect of feature consistency in image-augmentation. As depicted in Fig. \ref{fig2:Net} (c), our feature knowledge alignment strategy is designed to alleviate this problem. It prompts the features of weakly augmented view to supervise those of strongly augmented one through two primary perspectives: point-to-point alignment and prototype-based intra-class feature compactness.

\textbf{Point-to-point alignment.}  
The feature knowledge alignment, focusing on point-to-point alignment, is designed to preserve consistent contextual information across feature representations $F_s$ and $F_w$ of differently augmented views. 

In alignment with prior studies such as UniMatch \cite{yang2023revisiting} and AugSeg \cite{zhao2023augmentation}, we apply both strong and weak augmentations ($A_s(\cdot)$ and $A_w(\cdot)$) \footnote{The strong and weak augmentation methods utilized in this
study are adapted from UniMatch \cite{yang2023revisiting}.
} to the unlabeled image $u$ , resulting in a strongly augmented view $u_s$ and a weakly augmented view $u_w$:
\begin{equation}
    {u_s} = {A_s}(u)\;,\quad {u_w} = {A_w}(u).
    \label{eq1}
\end{equation}
$u_s$ and $u_w$ are fed into the encoder $f_t(\cdot)$ to obtain the corresponding feature representations $F_s$ and $F_w$ :
\begin{equation}
    {F_s} = {f_t}({u_s})\;,\;\;{F_w} = {f_t}({u_w}).
  \label{eq2}
\end{equation}
Next, we compute the similarity $S_{p2p}$ between $F_s$  and $F_w$ according to the following equation:
\begin{equation}
    {S_{p2p}} = \frac{1}{{p{\rm{ \times }}q}}\sum\limits_{i = 1}^{p{\rm{ \times }}q} {\frac{{d_s^{i} \cdot d_w^{i}}}{{||d_s^{i}||{\rm{ \times ||}}d_w^{i}{\rm{||}}}}} .
     \label{eq3}
\end{equation}
$p{\rm{ \times }}q$ is the resolution of the $F_s$  and $F_w$. $d_s^{i}$ and $d_w^{i}$ denotes the point feature at the $i$-th position of the  $F_s$ and $F_w$ in  current iteration. 
The point-to-point alignment loss $\mathcal{L}_{p2p}$ is computed using the formula below:
\begin{equation}
   {\mathcal{L}_{p2p}} = 1 - \frac{1}{{{B_u}}}\sum\limits_{i = 1}^{{B_u}} {S_{p2p}^i},
    \label{eq4}
\end{equation}
where $B_u$ represents the batch size.

\vspace{0.1cm}
\textbf{Intra-class feature compactness.} The outlier features have always been ignored by  consistency regularization-based methods. The feature knowledge alignment strategy from the perspective of prototype-based intra-class feature compactness make the network pay more attention to the outlier features.

After obtaining $F_w$ with  Eq. (\ref{eq2}), we input it into the decoder to get the class probability distribution $p_w$:
\begin{equation}
      p_w = g_{t}(F_w).
      \label{eq5}
\end{equation}
Then we calculate the  feature set of class $k$ in weakly augmented view features as follows:
\begin{equation}
    {W_k} = {F_w} \cdot \mathcal{I}(\arg \max ({p_w}) = k).
    \label{eq6}
\end{equation}
The size of $W_k$ is $N_k$ × $C$, and $N_k$ is the number of features. $C$ is the number of channels. $\mathcal{I}(\cdot)$ is a judgment function. Similarly, we get the  features of class $k$ in  strongly augmented view using the following formula:
\begin{equation}
    {S_k} = {F_s} \cdot \mathcal{I}(\arg \max ({p_w}) = k).
    \label{eq7}
\end{equation}
Next, we filter the intra-cluster features $M_{in}^k$ in the weakly augmented view features:
\begin{equation}
    {M_{in}^k} = \{ {r_k}|\cos ({r_k},{\rho _k}) \ge a\},
    \label{eq8}
\end{equation}
where $r_k$ is selected from $W_k$. We rank the features in $W_k$  based on the cosine similarity $\cos(\cdot)$ between the features of class $k$ in the current mini-batch and the class prototype $\rho _k$, and $a$ denotes the $N_r$-th high cosine similarity value. On the other hand, we select the outlier features $M_{dis}^k$ in  strongly augmented view features as follows:
\begin{equation}
    {M_{dis}^k} = \{ {h_k}|\cos ({h_k},{\rho _k}) \le  e\},
    \label{eq9}
\end{equation}
where $h_k$ is selected from $S_k$. Similar to Eq. (\ref{eq8}), $e$ is the $N_d$-th lowest cosine similarity value. Finally, we compute the outlier feature loss $\mathcal{L}_{dt}$ using the following formula:
\begin{equation}
    {\mathcal{L}_{dt}} = \frac{1}{Z}\sum\limits_{i = 1}^Z {\frac{1}{{{N_d}}}\sum {Near\_\cos (M_{in}^i,M_{dis}^i)} }, 
    \label{eq10}
\end{equation}
where $Near\_\cos(\cdot)$  is committed to finding the nearest intra-cluster feature $r_i$ for each outlier feature $h_i$ in $M_{dis}^k$, and utilizes the similarity-based loss function to align feature $h_i$ with feature $r_i$. $Z$ is the number of classes in the dataset.

In addition to this, we need to update the prototype of class $k$  to filter out more reliable intra-cluster features, so that the network can more accurately focus on outlier features:
\begin{equation}
{\rho '_{k}} = \eta {\rho _k} + (1 - \eta )\frac{1}{{{N_k}}}\sum\limits_{i = 1}^{{N_k}} {W_k^i},
    \label{eq11}
\end{equation}
where $\eta$ is a hyper-parameter that controls the updating speed. Following PCR \cite{xu2022semi}, we set $\eta$ to 0.99.

 \begin{figure}[t]
 \centering
 \includegraphics[width=\linewidth]{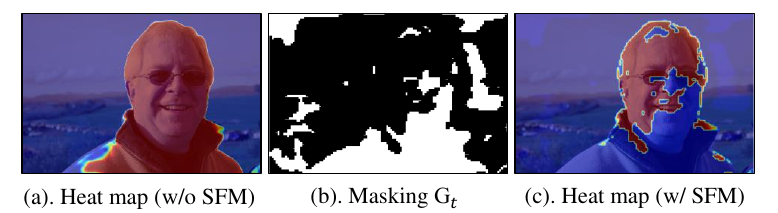}
 \caption{The visualization of self-adaptive feature masking (SFM). Regions that appear bluer receive less activation, while those that are redder receive higher activation. White regions in the  $G_t$ represent areas where activation needs to be filtered.}
 \label{fig3:heatmap}
 \end{figure}

\subsection{Self-adaptive Intervention}
Our feature knowledge alignment strategy motivates the encoder to attain feature consistency through image-augmentation. However, it inadvertently restricts the learning scope of the decoder. To address this, as depicted in Fig. \ref{fig2:Net} (b) and drawing inspiration from the feature perturbations in CCT \cite{ouali2020semi}, we introduce a self-adaptive intervention module. This module is designed to enhance the prediction consistency of the decoder via feature perturbation, thereby broadening the decoder's capacity to assimilate valuable information within the feature space. The module encompasses two main components: self-adaptive feature masking and noise injection.

\begin{algorithm}[t]
	\caption{Pseudocode of MCCL in a Pytorch-like style.}
	\label{alg:1}
	% \textcolor{white}{} \\
	\textcolor{white}{} \\
	\textcolor{mycolor1}{\# \  $f_t$: encoder, \ $g_t$: decoder.   $U(\cdot)$: uniform distribution.} \\
	\textcolor{mycolor1}{\# \ $\rho _k$: prototype of class $k$, \  C: the number of classes.} \\
	\textcolor{mycolor1}{\# \  $MSE$: mean squared error, ($\alpha, \omega, \beta$): loss weight. } \\
	\textcolor{mycolor1}{\# \  weakly and strongly augmented views ($u_w$ and $u_s$)} \\
	\textcolor{mycolor1}{\# \  intervention, left and right boundary values ($v_u$, $b_l$ and $b_r$).} \\
	\textcolor{mycolor1}{\# \  $Near\_cos(\cdot)$: Nearest neighbor similarity loss.} \\
	\\
	\textcolor{red}{for} \ ($u_w$ , $u_s$) \  \textcolor{red}{in} \  loader\_u :\\
	% \textcolor{mycolor}{\quad \quad \  \# get the predictions of the view} \\
	\textcolor{mycolor2}{\quad \quad \  $F_w$, \ $F_s$ \ = \  $f_t$($u_w$), \  $f_t(u_s)$} \\
	\textcolor{mycolor2}{\quad \quad \  $p_w,\  p_s \  = \ g_t(F_w), \  g_t(F_s)$} \\
	\\
	\textcolor{mycolor1}{\quad \quad \  \# point-to-point alignment (IF consistency)} \\
	\textcolor{mycolor2}{\quad \quad \ cosine \  = \  nn.CosineSimilarity () } \\
	\textcolor{mycolor2}{\quad \quad \ $\mathcal{L}_{p2p}$ \ = \ 1 $-$  \ cosine ($F_w$,$F_s$) . mean ()} \\
	\\
	\textcolor{mycolor1}{\quad \quad  \  \# intra-class feature compactness (IF consistency)} \\
	\textcolor{mycolor2}{\quad \quad  \  $\mathcal{L}_{dt}$ \  = \ 0} \\
	\textcolor{red}{\quad \quad \ for} \ $k$ \  \textcolor{red}{in} \  range (C) :\\
	% \textcolor{mycolor}{\quad \quad \quad \quad \  \# intra-cluster features \& outlier  features} \\
	\textcolor{mycolor2}{\quad \quad \quad \quad \  indice =  torch.sort ( cosine ($F_w[p_w==k]$, $\rho _k$))[1]} \\
	\textcolor{mycolor2}{\quad \quad \quad \quad \  $M_{in}^k$ = $F_w$ [indice] [ $-$ $N_r$ : \ ] }\\
	\textcolor{mycolor2}{\quad \quad \quad \quad \  index =  torch.sort ( cosine ($F_s[p_w==k]$, $\rho _k$))[1]} \\
	\textcolor{mycolor2}{\quad \quad \quad \quad \  $M_{dis}^k$ = $F_s$ [index] [0 : $N_d$] }\\
	\textcolor{mycolor1}{\quad \quad \quad \quad \  \#  outlier  feature loss} \\
	\textcolor{mycolor2}{\quad \quad \quad \quad \  $\mathcal{L}_{dt}$ \  = \ $\mathcal{L}_{dt}$ \  + $Near\_cos$ ( $M_{in}^k$, $M_{dis}^k$) }\\
	\textcolor{mycolor2}{\quad \quad  \   $\mathcal{L}_{dt}$ \  = \  $\mathcal{L}_{dt}$ . mean ()} \\
	\\
	\textcolor{mycolor1}{\quad \quad  \  \# self-adaptive intervention (FP \& IP consistency)} \\
	\textcolor{mycolor2}{\quad \quad  \   $F_{mk} = F_s [avg (F_s) < max (avg(F_s)) \cdot U(b_l, b_r)] $} \\ 
	\textcolor{mycolor2}{\quad \quad  \   $F_{ne} = F_s \cdot U(-v_u,v_u) + F_s $} \\
	\textcolor{mycolor2}{\quad \quad  \   $\mathcal{L}_{m}, \ \mathcal{L}_{n} \  = MSE (p_w, g_t(F_{mk})), \   MSE(p_w, g_t(F_{ne}))$} \\
	\textcolor{mycolor2}{\quad \quad  \   $\mathcal{L}_u = {\alpha \mathcal{L}_{p2p}} + {\omega  \mathcal{L}_{dt}} + \beta ({\mathcal{L}_m} + {\mathcal{L}_n})$} \\
	\vspace{-0.2cm}
\end{algorithm}

\textbf{Self-adaptive feature masking.}  
%\textcolor{red}{The following sentences sound a bit abrupt and lack coherence.}
During training, the model frequently overemphasizes activated regions, adversely impacting its generalization performance. To mitigate this, adaptive masking of features is essential for improving the model's robustness, as shown in Fig. \ref{fig3:heatmap}. We commence by computing the self-adaptive intervention value $v_u$ using the point-to-point similarity $S_{p2p}$:
\begin{equation}
    {v_u} = \lambda {\rm{(1 + }}{{S}_{p2p}}{\rm{)}},
    \label{eq12}
\end{equation}
where $S_{p2p}$ $\in$ [-1,1], and $\lambda$ is the scaling factor. Eq. (\ref{eq12}) can be seen as adaptively adjusting the intervention value $v_u$ within a certain range based on the similarity ${S}_{p2p}$. The higher the similarity, the greater the intervention value. We obtain the self-adaptive intervention left and right boundary values ($b_l$ and $b_r$) according to the following equations:
\begin{equation}
    {b_l} = \max (0,\;\frac{9}{{10}} - v_u),
     \label{eq13}
\end{equation}
\begin{equation}
    b{}_r = \min (1,\;\frac{{11}}{{10}} - v_u).
     \label{eq14}
\end{equation}
The  masking matrix $G_t$ is then obtained as follows:
\begin{equation}
    {G_t} = \mathcal{I}(avg({F_s}) < \max (avg({F_s})) \cdot U({b_l},{b_r})),
     \label{eq15}
\end{equation}
where the  size of $G_t$ 
is $p{\rm{ \times }}q{\rm{ \times }}1$, and $avg(\cdot)$ denotes the average over the channel dimension. $U(\cdot)$  represents uniform distribution. Eqs. (\ref{eq13}) and (\ref{eq14}) are focused on determining the left and right boundary values of the intervention based on the $v_u$. As the intervention value increases, the corresponding left and right boundary values decrease, leading to more regions being filtered out as described in Eq. (\ref{eq15}). Thus, the core of self-adaptive feature masking lies in adaptively filtering out highly activated regions based on the similarity between strongly and weakly augmented view features.

 We perform Hadamard product $\odot$ between $F_s$ and $G_t$ to generate self-adaptive masking feature representations $F_{mk}$:
\begin{equation}
  {F_{mk}} = F_s \odot {G_t}.
   \label{eq16}
\end{equation}
 We then feed $F_{mk}$ into the decoder $g_t(\cdot)$ to obtain its prediction:
\begin{equation}
    {p_m} = {g_t}({F_{mk}}).
    \label{eq17}
\end{equation}
The self-adaptive feature masking loss $\mathcal{L}_m$ is defined as follows:
\begin{equation}
    {\mathcal{L}_m} = \frac{1}{{{B_u}}}\sum\limits_{i = 1}^{{B_u}} {d(p_m^i,p_w^i)},
     \label{eq18}
\end{equation}
where $d(\cdot)$ is a distance measure between two class probability distributions, and this paper details the impact of using three probability distribution measurement ways on model performance in the subsequent experiment section: KL divergence, cross-entropy (CE), and mean squared error (MSE).

\begin{table*}[t]
	\caption{Comparison of the performance with other state-of-the-art methods on the \textbf{original} Pascal VOC2012 dataset. 1/n represents the proportion of labeled data, and the remaining data are unlabeled.
	}
	\renewcommand\arraystretch{1.2}
	\vspace{-0.25cm}
	\label{table1}
	\begin{center}
		\setlength{\tabcolsep}{5.2mm}{
			\begin{tabular}{r|c|c|c|c|c|c}
				\toprule
				\multirow{2}{*}{\textbf{Methods~~~~~}}  & \multirow{2}{*}{\textbf{Publication}} & \multirow{2}{*}{\textbf{Backbone}}  & 
				\multicolumn{4}{c}{\textbf{Performances (\%)}} \\  
				\cmidrule(lr){4-7} & & & \textbf{1/8 (183)} 	& \textbf{ 1/4 (366)}    & \textbf{1/2 (732)}  & \textbf{Full (1464)} \\ 
				\midrule
				Supervised &  - & ResNet-50 & 52.26 &   61.65 &   66.72  &  72.94 \\
				Supervised &  - & Transformer & 63.62 &   70.76 &   75.44  &  77.01 \\
				ECS \cite{mendel2020semi} & ECCV 2020 & ResNet-50 & 70.20 & 72.60 & - & 76.30 \\
				PseudoSeg \cite{zou2021pseudoseg} & ICLR 2021  & ResNet-50 & 61.88  &  64.85 &  
				70.42 & {  71.00}\\
				CPS \cite{chen2021semi} & CVPR 2021 & ResNet-50 & 67.42 &  71.71 &  75.88 &  - \\
				PC$^2$Seg \cite{zhong2021pixel} & ICCV 2021 & ResNet-50 & 64.63 & 67.62  & 70.90 & 72.26 \\
				DCC \cite{Lai_2021_CVPR} & CVPR 2021 & ResNet-50 & 72.40 & 74.00 & - & 76.50 \\
				ELN \cite{kwon2022semi}  & CVPR 2022 & ResNet-50 & 73.20 &  74.63 & - & -\\
				GuidedMix-Net \cite{tu2022guidedmix} & AAAI 2022 & ResNet-50 & 73.40 & 75.50 & 76.50 & - \\
				MKD \cite{yuan2023} & ACMMM 2023 & ResNet-50 & 66.74 &   71.01  &  72.73  &   78.14 \\
				CPCL \cite{fan2023conservative} & TIP 2023 & ResNet-50 & 67.02 & 72.14 & 74.25 & - \\
				AugSeg \cite{zhao2023augmentation} & CVPR 2023 & ResNet-50 & 72.17 &   76.17  &   77.40 &  78.82  \\
				UniMatch \cite{yang2023revisiting} & CVPR 2023 & ResNet-50 & 72.48  &  75.96  &  77.39 &  78.70 \\
				SemiCVT \cite{Huang_2023_CVPR} & CVPR 2023 & Transformer & 71.26 & 74.99 & 78.54 & 80.32 \\
				ESL \cite{Ma_2023_ICCV} & ICCV 2023  & ResNet-50 & 69.50 &  72.63  &   74.69  &   77.11 \\ 
				\midrule
				\rowcolor{mycolor} \textbf{Ours~~~~}  & - & ResNet-50 & \textbf{75.22} & \textbf{76.45} &	\textbf{78.51} &\textbf{79.62} \\
				\rowcolor{mycolor} \textbf{Ours~~~~}  & - & Transformer & \textbf{75.91} & \textbf{79.89} &	\textbf{80.72} & \textbf{81.96} \\
				\bottomrule
		\end{tabular}}
	\end{center}
	\vspace{-0.4cm}
\end{table*}

\textbf{Self-adaptive feature noise injection.} 
In addition to feature masking, we propose the adaptive injection of noise into the features of strongly augmented views to enhance feature diversity. Adaptive generation of uniform noise $N_l$ is based on the intervention value $v_u$ as defined in Eq. (\ref{eq12}):
\begin{equation}
    {N_l} = R({\rm{ - }}{{v}_u},{v_u}),
     \label{eq19}
\end{equation}
where $R(\cdot)$ produces a tensor of the same size as $F_s$, with all positions taking values in $[-v_u,v_u]$. We then get the self-adaptive noise feature representations $F_{ne}$ by injecting uniform noise into feature representations of the strongly augmented view $F_s$:
\begin{equation}
    {F_{ne}} = {F_s} \odot {N_l} + {F_s}.
     \label{eq20}
\end{equation}
$F_{ne}$ is fed into the decoder $g_t(\cdot)$ to get the prediction $p_n$:
\begin{equation}
    {p_n} = {g_t}({F_{ne}}).
    \label{eq21}
\end{equation}
Finally, we compute the self-adaptive feature noise injection loss $\mathcal{L}_n$ according to the following equation:
\begin{equation}
    {\mathcal{L}_n} = \frac{1}{{{B_u}}}\sum\limits_{i = 1}^{{B_u}} {d(p_n^i,p_w^i)}.
     \label{eq22}
\end{equation}
Self-adaptive feature noise injection can assist the decoder in expanding the learning range in feature space and thus learning more useful information. 

The main difference between feature perturbations in CCT \cite{ouali2020semi} and ours is that our proposed self-adaptive intervention module intervenes on strongly augmented view features in an instance-specific manner based on the similarity of the differently augmented view features, while CCT perturbs features in an instance-independent manner. Furthermore, unlike CCT's feature perturbations that necessitate multiple decoders, our module efficiently operates with a single decoder, significantly reducing the complexity of the model.

\subsection{Overall Training Objective}
The detailed process of the proposed MCCL can be found in Algorithm \ref{alg:1}. We train the  network jointly with supervised loss $\mathcal{L}_s$ for labeled data and consistency loss $\mathcal{L}_u$ for unlabeled data. The overall training loss during each iteration is formulated as:
\begin{equation}
     \mathcal{L} =  \mathcal{L}_s + \mathcal{L}_u.
     \label{eq23}
\end{equation}
We apply the weak augmentation to the labeled data and then input them into the  network to get the predicted results:
\begin{equation}
    {\hat{y}} = {g_t}({f_t}({A_w}({x}))),
    \label{eq24}
\end{equation}
where $\hat{y}$ is the predicted class probability distribution score. The standard cross-entropy loss $\ell_{ce}(\cdot)$ is used to supervise network training on labeled data:
\begin{equation}
   \mathcal{L}_s =  {\frac{1}{B_{u}}\cdot \frac{1}{{H{\rm{ \times }}W}}\sum_{i=1}^{B_{u}}\sum\limits_{j = 1}^{H{\rm{ \times }}W}  {{{\ell}_{ce}}(\hat{y}^{i}(j),{y}^{i}(j))}},
   \label{eq25}
\end{equation}
where $H{\rm{ \times }}W$ is the resolution of the image. $j$ represents the $j$-th pixel on the image, and $y$ denotes the ground-truth.

The consistency loss $\mathcal{L}_u$  consists of the point-to-point alignment loss $\mathcal{L}_{p2p}$, the outlier feature loss $\mathcal{L}_{dt}$,  the self-adaptive feature masking loss $\mathcal{L}_m$ and the self-adaptive feature noise injection loss $\mathcal{L}_n$. The formula for consistency loss $\mathcal{L}_u$ is as follows:
\begin{equation}
        \mathcal{L}_u = {\alpha \mathcal{L}_{p2p}} + {\omega  \mathcal{L}_{dt}} + \beta ({\mathcal{L}_m} + {\mathcal{L}_n})
        \label{eq26},
\end{equation}
where $ \alpha$, $\omega $, and $ \beta$ are the weights of corresponding losses.

\begin{table*}[t]
	\caption{Comparison of the performance with other state-of-the-art methods on the \textbf{blended} Pascal VOC2012 dataset. 1/n represents the proportion of labeled data, and the remaining data are unlabeled. 
	}
	\vspace{-0.25cm}
	\renewcommand\arraystretch{1.2}
	\label{table2}
	\begin{center}
		\setlength{\tabcolsep}{4.3mm}{
			\begin{tabular}{r|c|c|c|c|c|c|c}
				\toprule
				\multirow{2}{*}{\textbf{Methods~~~~~}}  &  \multirow{2}{*}{\textbf{Publication}}   & 
				\multicolumn{3}{c|}{\textbf{ResNet-50}} & \multicolumn{3}{c}{\textbf{ResNet-101}} \\  
				\cmidrule(lr){3-8}  & & \textbf{1/16 (662)} 	& \textbf{ 1/8 (1323)}    & \textbf{1/4 (2646)}  & \textbf{1/16 (662)} 	& \textbf{ 1/8 (1323)}    & \textbf{1/4 (2646)}\\ 
				\midrule
				Supervised & -  & 62.40 & 68.20 & 72.30 & 67.50 & 71.10 & 74.20 \\
				MT \cite{tarvainen2017mean} & NeurIPS 2017 &  66.80 & 70.80 & 73.20 & 70.60 & 73.20 & 76.60  \\
				CCT \cite{ouali2020semi} & CVPR 2020 &  65.21 & 70.90 & 73.40 & 67.94 & 73.00 & 76.17\\
				CPS \cite{chen2021semi} & CVPR 2021 &  71.98 & 73.67 & 74.90  & 74.48 & 76.44 & 77.68\\
				ST++ \cite{yang2022st++} & CVPR 2022 &  72.60 & 74.40 & 75.40 & 74.50 & 76.30 & 76.60\\
				ELN \cite{kwon2022semi} & CVPR 2022 &  70.50 & 73.20 & 74.60 & 72.50 & 75.10 & 76.60\\
				U2PL \cite{wang2022semi} & CVPR 2022 &  72.00 & 75.10 & 76.20 & 74.40 & 77.60 & 78.70\\
				PS-MT \cite{liu2022perturbed} & CVPR 2022 &  72.83 & 75.70 & 76.43 & 75.50 & 78.20 & 78.70\\
				AugSeg \cite{zhao2023augmentation} & CVPR 2023 &  74.66 & 75.99 & 77.16 & 77.01 & 77.31 & 78.82\\
				CCVC \cite{wang2023conflict} & CVPR 2023 &  74.50 & 76.10 & 76.40 & 77.20 & 78.40 & 79.00\\
				UniMatch \cite{yang2023revisiting} & CVPR 2023 &  75.80  &  76.90  &  76.80 & 78.10 & 78.40 & 79.20\\
				ESL \cite{Ma_2023_ICCV} & ICCV 2023 &  73.41 & 75.86 & 76.80 & 76.36 & 78.57 & 79.02\\
				\midrule 
				\rowcolor{mycolor} \textbf{Ours~~~~~~}  & - &  \textbf{76.06} & \textbf{77.55} &	\textbf{77.27} & \textbf{78.53} & \textbf{78.71} & \textbf{79.30}\\
				\midrule 
				\midrule 
				\multirow{2}{*}{\textbf{Methods~~~~~}}  &  \multirow{2}{*}{\textbf{Publication}}   & 
				\multicolumn{6}{c}{\textbf{Transformer}}  \\ 
				\cmidrule(lr){3-8}  & & \multicolumn{2}{c|}{\textbf{1/16 (662)}} 	& \multicolumn{2}{c|}{\textbf{ 1/8 (1323)}}    & \multicolumn{2}{c}{\textbf{1/4 (2646)}} \\ 
				\midrule 
				Supervised & - & \multicolumn{2}{c|}{72.01} & \multicolumn{2}{c|}{73.20} & \multicolumn{2}{c}{76.62} \\
				SemiCVT \cite{Huang_2023_CVPR} & CVPR 2023 & \multicolumn{2}{c|}{78.20} & \multicolumn{2}{c|}{79.95} & \multicolumn{2}{c}{80.20} \\
				\rowcolor{mycolor} \textbf{Ours~~~~~~}  & - &  \multicolumn{2}{c|}{\textbf{80.82}} & \multicolumn{2}{c|}{\textbf{81.17}} &	\multicolumn{2}{c|}{\textbf{81.44}} \\
				
				\bottomrule
		\end{tabular}}
	\end{center}
	
\end{table*}

\begin{table*}[h]
	\caption{Comparison of the performance with other state-of-the-art methods on the Cityscapes dataset. 1/n represents the proportion of labeled data, and the remaining data are unlabeled.
	}
	\vspace{-0.25cm}
	\renewcommand\arraystretch{1.2}
	\label{table3}
	\begin{center}
		\setlength{\tabcolsep}{5.2mm}{
			\begin{tabular}{r|c|c|c|c|c|c}
				\toprule
				\multirow{2}{*}{\textbf{Methods~~~~~}}  & \multirow{2}{*}{\textbf{Publication}} & \multirow{2}{*}{\textbf{Backbone}}  & 
				\multicolumn{4}{c}{\textbf{Performances (\%)}} \\  
				\cmidrule(lr){4-7} & & & \textbf{1/16 (186)} 	& \textbf{ 1/8 (372)}    & \textbf{1/4 (744)} &  \textbf{1/2 (1488)} \\ 
				\midrule
				Supervised & - & ResNet-50 & 63.30 & 70.20 & 73.10 & 76.60 \\
				Supervised & - & Transformer & 67.17 & 73.10 & 75.12 & 78.55 \\
				MT \cite{tarvainen2017mean} & NeurIPS 2017 & ResNet-50 & 66.14 & 72.03 & 74.47 & 77.43\\
				CCT \cite{ouali2020semi} & CVPR 2020 & ResNet-50 & 66.35 & 72.46 & 75.68
				& 76.78 \\
				CPS \cite{chen2021semi} & CVPR 2021 & ResNet-50 & 69.79 & 74.39 & 76.85 & 78.64 \\
				%ST++ \cite{yang2022st++} & CVPR 2022 & ResNet-50 & - & 72.70 & 73.80 & - \\
				ELN \cite{kwon2022semi} & CVPR 2022 & ResNet-50 & - & 70.33 & 73.52 & 75.33 \\
				U2PL \cite{wang2022semi} & CVPR 2022 & ResNet-50 & 69.00 & 73.00 & 76.30 & 77.10 \\
				PS-MT \cite{liu2022perturbed} & CVPR 2022 & ResNet-50 & - & 75.76 & 76.92 & 77.64 \\
				% AugSeg \cite{zhao2023augmentation} & CVPR 2023 & ResNet-50 & 74.66 & 75.99 & 77.16 \\
				CPCL \cite{fan2023conservative} & TIP 2023 & ResNet-50 & 69.92 & 74.60 & 76.98 & 78.17 \\
				CCVC \cite{wang2023conflict} & CVPR 2023 & ResNet-50 & 74.90 & 76.40 & 77.30 & - \\
				UniMatch \cite{yang2023revisiting} & CVPR 2023 & ResNet-50 & 75.03  &  76.77  &  77.49 & 78.60 \\
				SemiCVT \cite{Huang_2023_CVPR} & CVPR 2023 & Transformer & 72.19 & 75.41 & 77.17 & 79.55 \\
				ESL \cite{Ma_2023_ICCV} & ICCV 2023 & ResNet-50 & 71.07 & 76.25 & 77.58 & 78.92 \\
				\midrule 
				\rowcolor{mycolor} \textbf{Ours~~~~}  & - & ResNet-50 & \textbf{75.69} & \textbf{76.83} &	\textbf{78.20} & \textbf{79.33}\\
				\rowcolor{mycolor} \textbf{Ours~~~~}  & - & Transformer & \textbf{74.12} & \textbf{77.03} &\textbf{78.63} & \textbf{80.87}\\
				\bottomrule
		\end{tabular}}
	\end{center}
	
\end{table*}

\section{EXPERIMENTS}
This section outlines the experiments of the proposed method on various datasets and detailed ablation studies. We first elaborate on the experimental datasets and demonstrate the details of our approach during training. Subsequently, we compare the performance of our method with other methods on multiple datasets. Finally, we conduct detailed ablation experiments on each component of the proposed method to demonstrate their effectiveness. Comprehensive analyses are carried out on both the comparative experimental results and the ablation study results.

\subsection{Experiment Setup}
\textbf{Datasets.} We evaluate the performance of our method on  Pascal VOC2012 \cite{everingham2015pascal} and Cityscapes \cite{cordts2016cityscapes} datasets. The  Pascal VOC2012  is a standard semantic segmentation dataset comprising  20 foreground classes and 1 background class. 1464 fine-labeled training images and 1449 validation images collectively constitute the \textbf{\textit{original}} Pascal VOC2012 dataset. Following the dataset setup of other state-of-the-art methods \cite{Ma_2023_ICCV,wang2023conflict}, we also adopt the \textbf{\textit{blended}} version of the  Pascal VOC2012 dataset, comprising a total of 10,582 images obtained by augmenting 9,118 coarsely-labeled images from the SBD \cite{hariharan2011semantic} dataset.
Cityscapes is an urban scenes dataset comprising 19 classes. It consists of 2,975 images for training, 500 images for validation, and 1,525 images for testing. 

\textbf{Evaluation Metric.} We conduct extensive comparative experiments on the \textbf{\textit{original}}, \textbf{\textit{blended}} Pascal VOC2012 and  Cityscapes datasets using the same proportion of labeled images as previous methods \cite{yang2023revisiting,zhao2023augmentation}. It is worth mentioning here that the Full (1464) setting indicates 1464 labeled images and 9118 unlabeled images. Mean Intersection over Union (mIoU) is employed as the evaluation metric.

\textbf{Implementation Details.} Our model code is built upon the PyTorch framework and trained on eight NVIDIA A6000 GPUs with  48 GB memory per card. We conduct experiments using two segmentation networks: the first is DeepLabv3+ \cite{chen2018encoder} with a ResNet-50/ResNet-101 \cite{he2016deep} backbone pretrained on ImageNet \cite{deng2009imagenet}, and the second is SegFormer-B5 \cite{xie2021segformer}, which is based on the Transformer architecture.
As with other state-of-the-art method \cite{yang2023revisiting}, for \textit{\textbf{original}} Pascal VOC2012, we set the input image size to 321$\times$321, and for \textit{\textbf{blended}} Pascal VOC2012, we set the input image size to 513$\times$513. The learning rate is consistently set to 0.001. In the case of the Cityscapes dataset, we use a learning rate of 0.005 and adjust the image size to 801$\times$801. The number of training epochs for the model is 80  and 240, respectively. The batch size for both datasets is set to 8,  which means each mini-batch consisted of 8 labeled images and 8 unlabeled images. Weak augmentation operations consist of horizontal and vertical flipping, while strong augmentation operations not only include weak augmentation operations, but also contain colorjitter, graying, and gaussian blur. 

%Cutmix \cite{yun2019cutmix},

\subsection{Comparison with State-of-the-Art Methods}

 \textbf{Results on \textit{original} Pascal VOC2012.} Table \ref{table1} presents the performance comparison of our approach with other state-of-the-art methods on the \textbf{\textit{original}} Pascal VOC2012 dataset. The term `Supervised' refers to training conducted solely with a specific proportion of labeled data, without incorporating any unlabeled data. As can be seen, our approach significantly outperforms the current state-of-the-art methods,  whether using ResNet or the Transformer as the backbone network. Specifically, with the ResNet backbone, our method exceeds UniMatch \cite{yang2023revisiting}  by 2.74\%, 0.49\%, 1.12\%, and 0.92\% mIoU in the 1/8, 1/4, 1/2, and Full settings, respectively. It also surpasses ESL \cite{Ma_2023_ICCV} by 5.72\%, 3.82\%, 3.82\%, and 2.51\% mIoU in these settings. Additionally, when employing the Transformer architecture, our method surpasses SemiCVT \cite{Huang_2023_CVPR} by 4.65\%, 4.90\%, 2.18\%, and 1.64\% mIoU in the 1/8, 1/4, 1/2, and Full settings, respectively. These results clearly demonstrate the effectiveness of our method, particularly in scenarios with extremely limited labeled data, such as the 1/8 setting.

\begin{table*}[t]
\caption{Ablation experiments of Image-augmentation based Prediction (IP) consistency, Image-augmentation based Feature (IF) consistency and Feature-perturbation based Prediction (FP) consistency  on \textbf{original} Pascal VOC2012 and Cityscapes datasets.}
\vspace{-0.45cm}
\renewcommand\arraystretch{1.2}
\label{table4}
\begin{center}
\setlength{\tabcolsep}{5mm}{
\begin{tabular}{cccc|c|cc|cc}
\toprule
\multirow{2}{*}{\textbf{Baseline}}  & \multirow{2}{*}{\textbf{IP}} & \multirow{2}{*}{\textbf{IF}}  & \multirow{2}{*}{\textbf{FP}} & \multirow{2}{*}{\textbf{Backbone}} &
\multicolumn{2}{c|}{\textbf{\textit{Original} Pascal VOC2012}} & \multicolumn{2}{c}{\textbf{Cityscapes}} \\  
\cmidrule(lr){6-9} 
& & & & & \textbf{ 1/2 (732)}    & \textbf{Full (1464)}  & \textbf{1/16 (186)} & \textbf{1/4 (744)} \\ 
\midrule
$\checkmark$ &   & & & ResNet-50 & 66.72 & 72.94 & 63.30 & 73.10 \\
$\checkmark$ & $\checkmark$  & &  & ResNet-50 & 77.39 & 78.70 & 75.03 & 77.49 \\
$\checkmark$ & $\checkmark$  & $\checkmark$ & & ResNet-50 & 78.33 & 79.36 & 75.47 & 77.99 \\
$\checkmark$ & $\checkmark$  &  & $\checkmark$ & ResNet-50 & 77.98 & 79.29 & 75.36 & 77.78 \\
$\checkmark$ &  $\checkmark$ & $\checkmark$ & $\checkmark$ & ResNet-50&\cellcolor{mycolor}\textbf{78.51} & \cellcolor{mycolor}\textbf{79.62} & \cellcolor{mycolor}\textbf{75.69} & \cellcolor{mycolor}\textbf{78.20}\\

\bottomrule
\end{tabular}}
\end{center}
\vspace{-0.2cm}
\end{table*}

\textbf{Results on \textit{blended} Pascal VOC2012.} Table \ref{table2} displays the performance comparison of our method with other state-of-the-art methods on the \textbf{\textit{blended}} Pascal VOC2012 dataset. The data presented in the table demonstrate that our proposed MCCL method considerably exceeds other approaches across various configurations, regardless of the encoder used—whether it is ResNet-50, ResNet-101, or a Transformer. Specifically, with ResNet-50 as the backbone network, our method outperforms ESL \cite{Ma_2023_ICCV} by 2.65\%, 1.69\%, and 0.47\% mIoU under the 1/16, 1/8, and 1/4 labeled data ratios, respectively, while also exceeding CCVC \cite{wang2023conflict} by 1.56\%, 1.45\%, and 0.87\% mIoU. Similarly, employing ResNet-101 as the encoder, our method outperforms AugSeg \cite{zhao2023augmentation} by 1.52\%, 1.4\%, and 0.48\% mIoU in the corresponding labeled data partition settings.
Our proposed approach surpasses the SemiCVT \cite{Huang_2023_CVPR} by 2.62\%, 1.22\%, and 1.24\% mIoU across various configurations, highlighting its effectiveness within the Transformer architecture. The data comparisons above all demonstrate the superior performance of our method.

\textbf{Results on Cityscapes.} Table \ref{table3} presents a comparative analysis of MCCL against other state-of-the-art methods on the Cityscapes dataset. As can be seen, our proposed MCCL outperforms all other methods in various settings. 
Especially when using a Transformer as the backbone network,  our method outperforms SemiCVT \cite{Huang_2023_CVPR} by 1.93\%, 1.62\%, 1.46\%, and 1.32\% mIoU across the 1/16, 1/8, 1/4, and 1/2 settings, respectively. Moreover, under the same conditions, our method surpasses UniMatch \cite{yang2023revisiting} by 0.7\% mIoU at the 1/2 setting.
A notable observation is that MCCL shows a more significant performance improvement on the Pascal VOC 2012 dataset compared to Cityscapes,  which could be attributed to the higher complexity of scene composition in Cityscapes. The performance results  show that when using the Transformer as the backbone network, it outperforms ResNet in the 1/8, 1/4, and 1/2 settings, but slightly underperforms in the 1/16 setting. We hypothesize that although the Transformer efficiently captures scene information, when the amount of labeled data is too small, the Transformer may overfit the data, preventing it from learning more effective global information.  Analyzing results from both Pascal VOC2012 and Cityscapes datasets, it is evident that compared to prior consistency regularization methods, our staged enhancement of the encoder and decoder, along with additional consistency constraints, sets a new benchmark in performance.

\begin{figure}[t]
\centering
\setlength{\abovecaptionskip}{-0.005cm}
\includegraphics[width=1\linewidth]{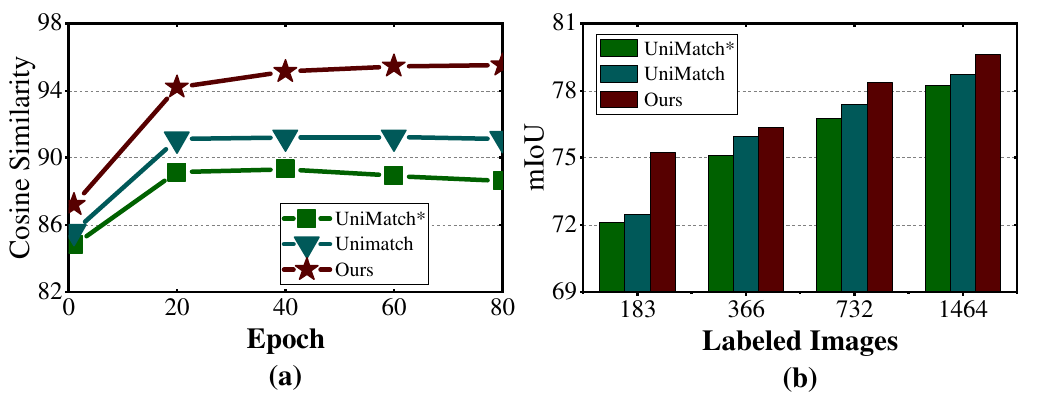}
\caption{Impact of Image-augmentation based Feature (IF) consistency on model performance using the \textbf{\textit{original}} Pascal VOC2012 dataset. (a) Similarity variation curve of strongly-weakly augmented view features under 1464 labeled image setting. (b) Comparison with other state-of-the-art methods under different labeled image settings. UniMatch$^*$ denotes a variant of UniMatch with dissimilar features. }
\label{fig4:mIou_similarity}
\vspace{-0.2cm}
\end{figure}

 \begin{figure*}[t]
\centering
\includegraphics[width=\linewidth]{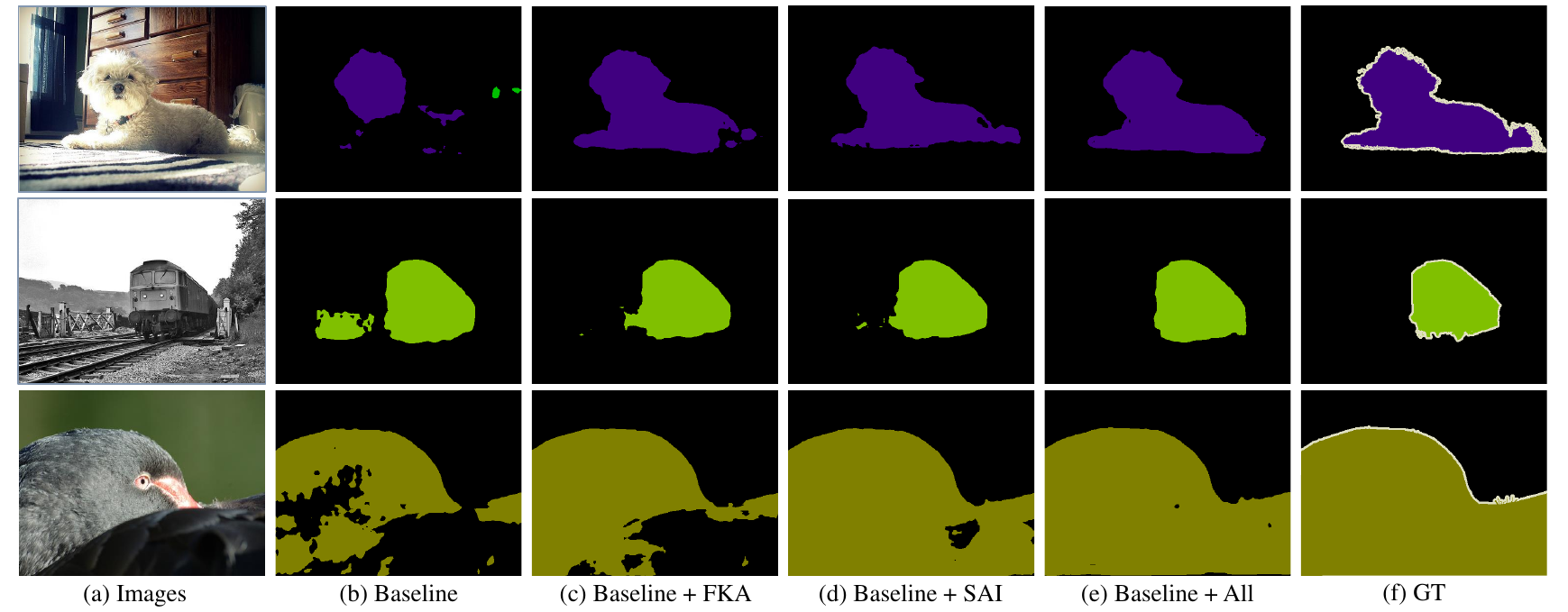}
\caption{Visual comparison of ablation results for FKA and SAI on the Pascal VOC2012 dataset. GT means Ground-Truth.} 
\label{fig5:Ablation_visual}
\end{figure*}

\subsection{Ablation Studies}
\indent This section delineates the comprehensive results of ablation studies conducted on each element of our method, thereby elucidating the contribution of individual components to the model's overall performance. All ablation experiments are conducted under a consistent experimental framework, employing ResNet-50 as the encoder and DeepLabv3+ as the decoder. The baseline used in all ablation experiments defaults to UniMatch \cite{yang2023revisiting}.

 \textbf{Impact of feature consistency.} We conduct detailed ablation experiments on the Image-augmentation based Feature (IF) consistency, and the experimental results are shown in Fig. \ref{fig4:mIou_similarity}. MCCL performs significantly better than the state-of-the-art method UniMatch \cite{yang2023revisiting} in feature consistency, outperforming  UniMatch significantly in different labeled data settings. Besides, we apply a loss to UniMatch that makes the features of strongly and weakly augmented views dissimilar, resulting in significant performance degradation of UniMatch$^*$. From the above experimental results, it can be seen that mining feature consistency information is beneficial for improving the performance of the model.

 \textbf{Impact of IP, IF and FP consistency.} We perform comprehensive ablation experiments on the \textbf{\textit{original}} Pascal VOC2012 and Cityscapes datasets, assessing Image-augmentation based Prediction (IP) consistency, Image-augmentation based Feature (IF) consistency, and Feature-perturbation Prediction (FP) consistency. The experimental results are detailed in Table \ref{table4}. The experimental results indicate that the exclusive application of IP consistency invariably enhances the model's performance. Specifically, under the 1/16 setting of the Cityscapes dataset, IP consistency has resulted in an 11.73\% increase in mIoU. When applied independently across different labeled data ratios, both IF and FP consistency deliver substantial performance gains. 
 We observe that the performance improvement from IF consistency is more pronounced than that from FP.  While FP consistency employs self-adaptive intervention on features, its mechanism of action is similar to that of IP consistency, with both approaches focusing on alignment at the segmentation level. In contrast, IF consistency emphasizes alignment at the feature level, thereby complementing the role of IP consistency and significantly enhancing overall model performance.
 Best experimental results are achieved when IP, IF, and FP are cohesively integrated into the baseline model.

\begin{figure}[t]
 \setlength{\abovecaptionskip}{-0.0001cm}
\centering
\includegraphics[width=\linewidth]{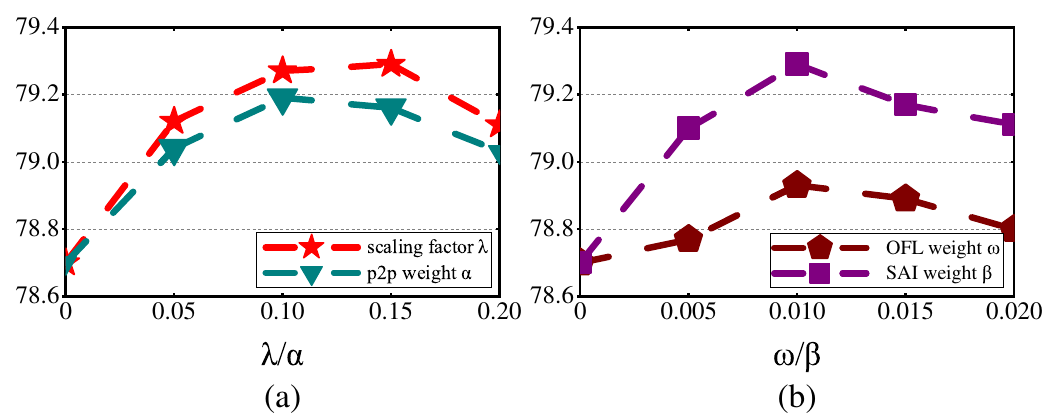}
\caption{Effect of hyper-parameters of MCCL on the \textbf{\textit{original}} Pascal VOC2012 dataset. (a). Ablations of scaling factor $\lambda$ and point-to-point (p2p) weight $\alpha$ on the model performance under the Full (1464) setting. (b). Ablations of self-adaptive intervention (SAI) loss weight  $\beta$ and the outlier feature loss (OFL) weight  $\omega$ on model performance under the Full setting.}

\label{fig6:ablations}
% \vspace{-0.18cm}
\end{figure}

  \textbf{Scaling factor.} We conduct ablation experiments on the scaling factor $\lambda$ (Eq. (\ref{eq12})) in the self-adaptive intervention module, and the results are presented in Fig. \ref{fig6:ablations}. The experimental results indicate that appropriately increasing the scaling factor to enhance intervention can significantly improve the performance of the model. Nonetheless, it is observed that surpassing a certain threshold in the scaling factor value leads to a gradual deterioration in performance. The best results are achieved when the scaling factor $\lambda$ = 0.15.

   \textbf{Loss weight.}  Fig. \ref{fig6:ablations} shows the impact of the weights ($\alpha$, $\omega$ and $\beta$) for point-to-point alignment loss, outlier feature loss, and self-adaptive intervention loss on model performance. The experimental results from the figure indicate that setting the weights within a specific range can significantly improve the performance of the model. When $\alpha$ = 0.1, $\beta$ = 0.01, and $\omega$ = 0.01, the model achieves the best performance.

\begin{table}[t]
\caption{Effect of each component in our method using the Full setting on the \textbf{original} Pascal VOC2012 dataset. FKA represents the feature knowledge alignment strategy, while SAI is the self-adaptive intervention module.}
\vspace{-0.25cm}
\renewcommand\arraystretch{1.2}
\label{table5}
\begin{center}
\setlength{\tabcolsep}{3mm}{
\begin{tabular}{ccccc|c}
\toprule
\multirow{2}{*}{\textbf{Baseline~~~~~}}  & \multicolumn{2}{|c|}{\textbf{FKA}} & \multicolumn{2}{|c|}{\textbf{SAI}} & \multirow{2}{*}{\textbf{mIoU}}\\
\cmidrule(lr){2-3}  \cmidrule(lr){4-5} 
& \multicolumn{1}{|c}{$ \mathcal{L}_{p2p} $} &  $ \mathcal{L}_{dt} $ & \multicolumn{1}{|c}{$ \mathcal{L}_n$} &  $ \mathcal{L}_m $ & \\
\midrule
$\checkmark$ &   & & & &  78.70\\
$\checkmark$ & $\checkmark$ & & & &  79.19 \\
$\checkmark$ &  & $\checkmark$ & & &  78.93 \\
$\checkmark$ & $\checkmark$ & $\checkmark$ & & &  79.36 \\
$\checkmark$ &  &  & $\checkmark$ & &  78.99 \\
$\checkmark$ &  &  &  & $\checkmark$ &  79.23 \\
$\checkmark$ &  &  & $\checkmark$ & $\checkmark$ &  79.29 \\
$\checkmark$ & $\checkmark$ & $\checkmark$ & $\checkmark$ & $\checkmark$ &  \textbf{79.62} \\
\bottomrule
\end{tabular}}
\end{center}
\end{table}

\begin{table}[t]
\caption{Comparison with conventional feature perturbations F-Noise (Feature-Noise) and F-Drop (Feature-Drop) from CCT using the Full setting (1464) on the \textbf{original} Pascal VOC2012 dataset.}
\vspace{-0.25cm}
\renewcommand\arraystretch{1.2}
\label{table6}
\begin{center}
\setlength{\tabcolsep}{1.6mm}{
\begin{tabular}{ccccc|c}
\toprule
\multirow{2}{*}{\textbf{Baseline~~~~~}}  & \multicolumn{2}{|c|}{\textbf{CCT}} & \multicolumn{2}{|c|}{\textbf{Ours}} & \multirow{2}{*}{\textbf{mIoU}}\\
\cmidrule(lr){2-3}  \cmidrule(lr){4-5} 
& \multicolumn{1}{|c}{ F-Noise} &   F-Drop  & \multicolumn{1}{|c}{SAF-Noise} &  SAF-Masking & \\
\midrule
$\checkmark$ &   & & & &  78.70\\
$\checkmark$ & $\checkmark$ & & & &  78.75 \\
$\checkmark$ &  & $\checkmark$ & & &  78.89 \\
$\checkmark$ &  &  & $\checkmark$ & &  78.99 \\
$\checkmark$ &  &  &  &$\checkmark$ &  \textbf{79.23} \\
\bottomrule
\end{tabular}}
\end{center}
\end{table}

\begin{table}[t]
\caption{Effect of $N_r$ for model performance using the 1/8 setting (183) on the \textbf{original} Pascal VOC2012 dataset.}
\vspace{-0.25cm}
\renewcommand\arraystretch{1.2}
\label{table7}
\begin{center}
\setlength{\tabcolsep}{1.6mm}{
\begin{tabular}{c|cccccc}
\toprule
$N_r$ & 0 & 8 & 12 & 16 & 20 & 24\\
\midrule
mIoU &  72.48 & 73.34 &  73.77 & \textbf{73.99} & 73.81 & 73.52 \\
\bottomrule
\end{tabular}}
\end{center}
\end{table}

\begin{table}[!htbp]
\caption{Effect of $N_d$ for model performance using the 1/8 setting (183) on the \textbf{original} Pascal VOC2012 dataset.}
\vspace{-0.25cm}
\renewcommand\arraystretch{1.2}
\label{table8}
\begin{center}
\setlength{\tabcolsep}{1.6mm}{
\begin{tabular}{c|cccccc}
\toprule
$N_d$ & 0 & 64 & 128 & 256 & 512 & 1024\\
\midrule
mIoU &  72.48 & 73.45 &  73.65 & \textbf{73.99} & 73.92 & 73.62 \\
\bottomrule
\end{tabular}}
\end{center}
\end{table}

\begin{table}[t]
\caption{Ablation experiments on the impact of probability distribution measurement ways on model performance using the 1/2 setting of the \textbf{original} Pascal VOC2012 dataset.}
\vspace{-0.25cm}
\renewcommand\arraystretch{1.2}
\label{table9}
\begin{center}
\setlength{\tabcolsep}{2mm}{
\begin{tabular}{cccc|c}
\toprule
Baseline  &  KL divergence & CE  & MSE & mIoU\\
\midrule
$\checkmark$ & & & & 77.39 \\
$\checkmark$ & $\checkmark$ & & & 77.92 \\
$\checkmark$ &  & $\checkmark$ & & 78.12 \\
$\checkmark$ &  &  & $\checkmark$ & \textbf{78.51} \\
\bottomrule
\end{tabular}}
\end{center}
\end{table}

\begin{figure*}[t]
	\centering
	\includegraphics[width=\linewidth]{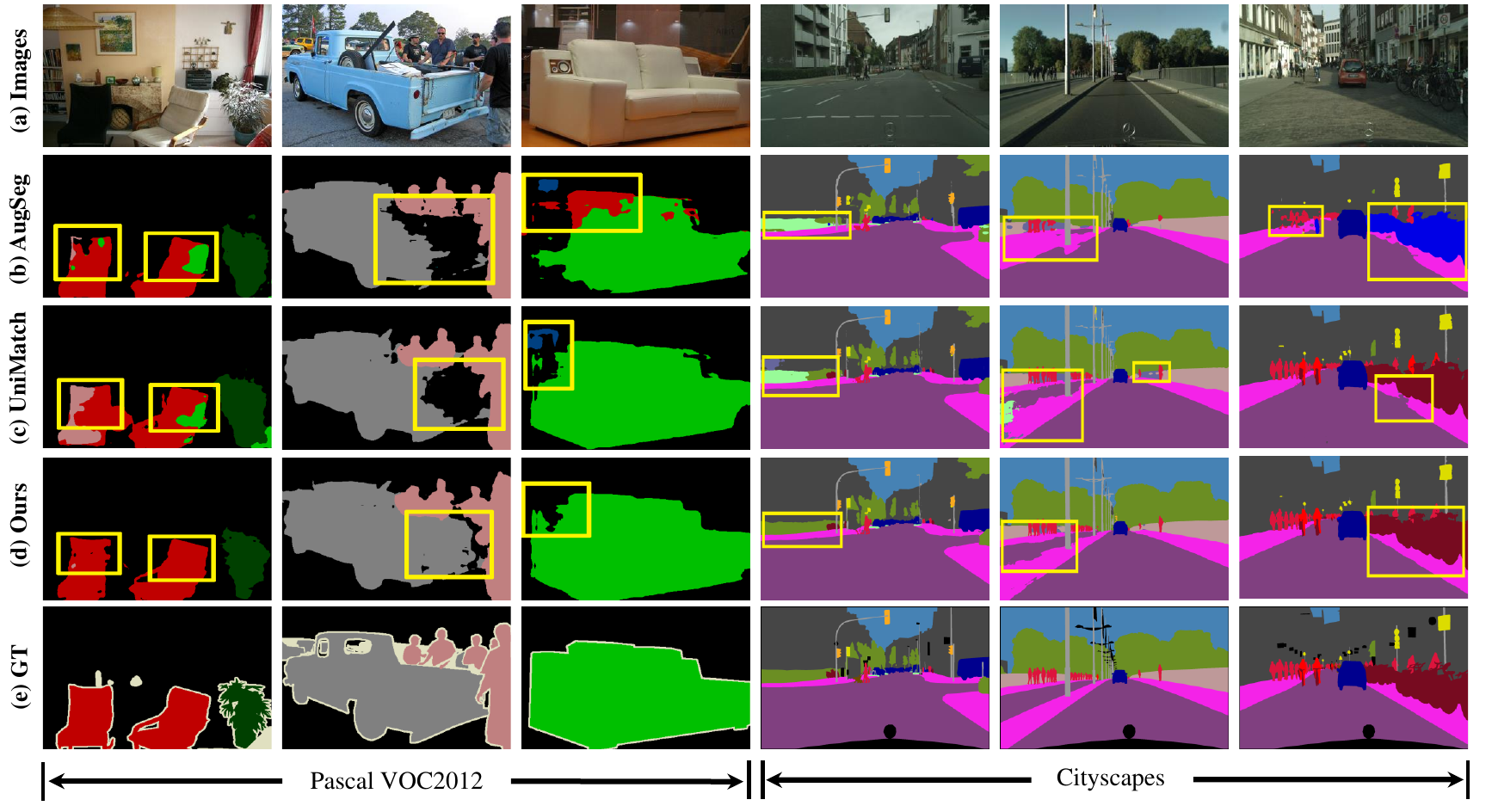}
	\caption{Comparison of visualization results on the Pascal VOC2012 and Cityscapes datasets. GT means Ground Truth.} 
	\label{fig7:visual}
\end{figure*}

\begin{figure}[t]
	\centering
	\includegraphics[width=\linewidth]{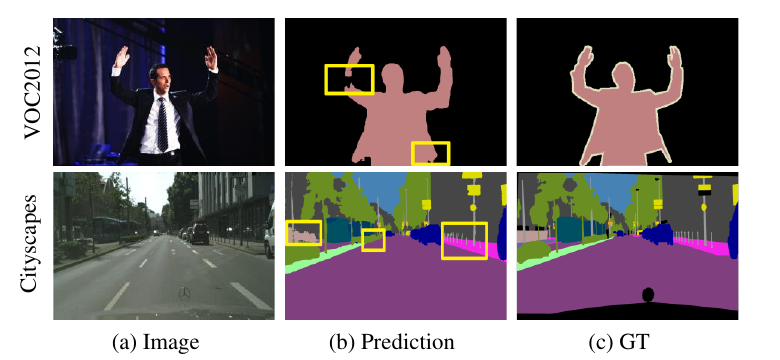}
	\caption{Failure cases on the Pascal VOC2012 and Cityscapes datasets.} 
	\label{fig8:failure}
\end{figure}

\textbf{Component analysis.} Table \ref{table5} shows the results of the ablation experiments for each component of our method. It can be seen that adding the feature knowledge alignment (FKA) strategy can improve the performance of the original baseline from 78.70\% to 79.36\% mIoU, demonstrating the effectiveness of feature consistency based on image-augmentation. Specifically, the point-to-point alignment loss $\mathcal{L}_{p2p}$ and the outlier feature loss $\mathcal{L}_{dt}$ improve the performance of the model by 0.49\% and 0.23\% mIoU, respectively. In addition, the self-adaptive intervention (SAI) module can yield a 0.59\% performance gain compared to the baseline. Self-adaptive noise injection loss $\mathcal{L}_n$ and self-adaptive feature masking loss $\mathcal{L}_m$ bring performance improvements of 0.29\% and 0.53\% mIoU to the model, respectively. 
This proves that applying self-adaptive intervention to the features of strongly augmented views can promote the decoder to learn more useful information from the broader feature space. In addition, we conduct a visual comparison of the segmentation results for FKA and SAI, as shown in Fig. \ref{fig5:Ablation_visual}.
Both qualitative and quantitative results indicate that the complementary integration of the feature knowledge alignment strategy and the self-adaptive intervention module can enable the model to achieve the best experimental performance.

\textbf{Self-adaptive \textit{vs} Conventional feature perturbation.} The performance comparison between the self-adaptive intervention module and conventional feature perturbations is shown in Table \ref{table6}. Compared with conventional feature perturbations, adopting a self-adaptive intervention module for strongly augmented view features can significantly improve the performance of the model. Specifically, compared to F-Noise which only improves the baseline with 0.05\% mIoU, our self-adaptive feature noise injection (SAF-Noise) can bring 0.29\% mIoU performance gain. On the other hand, when F-Drop can increase by 0.19\% mIoU, our self-adaptive feature masking (SAF-Masking) significantly outperforms the baseline by 0.53\% mIoU. The experimental results demonstrate that our self-adaptive intervention module, in an instance-specific manner, can promote the decoder to learn more useful information in broader feature space, thereby significantly improving the performance of the model.

\textbf{The impact of $N_r$ and $N_d$.} Tables \ref{table7} and \ref{table8} show the impact of the values of $N_r$ and $N_d$ on model performance.   Compared with the baseline ($N_r$/$N_d$=0), setting the values of $N_r$ and $N_d$ reasonably can greatly improve the model. 
The experimental results show that the model achieves the best performance when $N_r$ is set to 16 and $N_d$ is set to 256. In other words, the 256 outlier features are compacted to the 16 intra-cluster features by the nearest-neighbor similarity loss, making the model pay more attention to the outlier features during the training process.

\textbf{Probability distribution measurement analysis.}  We perform detailed ablation experiments on a variety of probability distribution measurement ways in Eqs. (\ref{eq18}) and (\ref{eq22}), with the results presented in Table \ref{table9}. These results unequivocally demonstrate that mean squared error (MSE) yields the most substantial improvement in model performance. we speculate that compared to KL divergence or cross-entropy (CE), MSE is more sensitive to smaller differences between distributions and does not need to consider whether the distribution satisfies the specific conditions, resulting in the best performance.

\subsection{Comparison of Visualization Results}
Fig. \ref{fig7:visual} displays the segmentation results on the Pascal VOC2012 and Cityscapes datasets using ResNet-50 as the encoder. It compares our method with other state-of-the-art methods (AugSeg \cite{zhao2023augmentation} and UniMatch \cite{yang2023revisiting}). The visualization results clearly show that our method precisely identifies target areas and their boundaries. For example, AugSeg and UniMatch cannot accurately segment the car area in the second column, and they incorrectly classify some pixels that are part of the car as background. In the third column, our method more accurately segments the boundaries of the sofa area compared to the other two methods. The segmentation results on the Cityscapes dataset demonstrate that our method maintains superior performance even in complex scenes. The results above fully show that our method can achieve state-of-the-art experimental performance.

\subsection{Failure Cases and Limitation}
Although our approach has achieved state-of-the-art experimental performance compared to previous methods, the segmentation results in Fig. \ref{fig8:failure} show that our approach still has some limitations in certain scenarios. The results in the first row demonstrate that our method struggles to segment regions of different categories that have similar appearances, \textit{e.g.}, it incorrectly classifies the background as a person. The second row reveals that elongated objects (poles) in the image are not accurately segmented.

\section{Conclusion}
In this paper, we present a Multi-Constraint Consistency Learning approach to facilitate the staged enhancement of the encoder and decoder by imposing additional consistency constraints on the network, thereby fully utilizing the potential supervisory information in the network.
Specifically, we first propose a feature knowledge alignment strategy based on image-augmentation from the perspectives of point-to-point alignment and prototype-based intra-class compactness to promote feature consistency learning of the encoder. In addition, we further propose a self-adaptive intervention module to encourage prediction consistency learning of the decoder based on feature-perturbation.
Extensive comparative and ablation experiments on the Pascal VOC2012 and Cityscapes datasets demonstrate that our method achieves state-of-the-art performance.

\section*{Acknowledgments}

This work was supported by the National Natural Science Foundation of China (No. 62472222, 62302217), Natural Science Foundation of Jiangsu Province (No. BK20240080, BK20220934, BK20220936), China Postdoctoral Science Foundation (No. 2022M711635), Jiangsu Funding Program for Excellent Postdoctoral Talent (No. 2022ZB267), Fundamental Research Funds for the Central Universities (No. 30923010303).

\bibliographystyle{IEEEtran}
\bibliography{ref}

\vfill

\end{document}